\documentclass[twocolumn]{aastex631}

\usepackage{amsmath,amssymb}
\usepackage{graphicx}
\usepackage[utf8]{inputenc}
\usepackage{dcolumn}
\usepackage{bm}
\usepackage{xcolor}
\usepackage{multirow}
\usepackage{array}
\newcolumntype{P}[1]{>{\centering\arraybackslash}p{#1}}

\def\barnue{\mathrel{{\bar \nu}_e}}

\def \sun	 	{$_{\odot}$}
\def\h{\hat{h}}
\def\d{\delta}

\def\positron{\mathrel{{e^+}}}

\def\t13{\mathrel{{\theta_{13}}}}
\def\y12{\mathrel{{\tan^2 \theta_{12}}}}
\def\c2{\mathrel{{\chi^2 }}}

\def \sun	 	{$_{\odot}$}

\newcommand{\n}{neutrino}
\newcommand{\ns}{neutrinos}
\newcommand{\sn}{supernova}

\newcommand{\be}{\begin{equation}}
\newcommand{\ee}{\end{equation}}
\newcommand{\ba}{\begin{eqnarray}}
\newcommand{\ea}{\end{eqnarray}}

\submitjournal{ApJ}
\shorttitle{Gamma ray echo of a supernova neutrino burst}
\shortauthors{Lunardini et al.}
\begin{document}
\title{Photons from neutrinos: the gamma ray echo of a supernova neutrino burst}

\author[0000-0002-9253-1663]{Cecilia Lunardini}
\email{Cecilia.Lunardini@asu.edu}
\affiliation{Department of Physics, Arizona State University, Tempe, AZ 85287-1504 USA}

\author{Joshua Loeffler}
\affiliation{Department of Physics, Arizona State University, Tempe, AZ 85287-1504 USA}

\author[0000-0002-2109-5315]{Mainak Mukhopadhyay}
\email{mkm7190@psu.edu}
\affiliation{Department of Physics; Department of Astronomy \& Astrophysics; Center for Multimessenger Astrophysics, Institute for Gravitation and the Cosmos, The Pennsylvania State University, University Park, PA 16802, USA}

\author[0000-0003-1513-3116]{Matthew J. Hurley}
\affiliation{Department of Physics, Stanford University, Stanford, California 94305, USA}

\author[0000-0002-5794-4286]{Ebraheem Farag}
\affiliation{School of Earth and Space Exploration, Arizona State University, Tempe, AZ 85287, USA.}
\affiliation{Department of Astronomy, Yale University, New Haven, CT 06511}

\author[0000-0002-0474-159X]{F.X.~Timmes}
\affiliation{School of Earth and Space Exploration, Arizona State University, Tempe, AZ 85287, USA.}

\begin{abstract}
When a star undergoes core collapse, a vast amount of energy is released in a $\sim 10$ s long burst of neutrinos of all species. Inverse beta decay in the star's hydrogen envelope causes an electromagnetic cascade which ultimately results in a flare of gamma rays -- an ``echo" of the neutrino burst --  at the characteristic energy of 0.511 MeV. We study the phenomenology and detectability of this flare. Its luminosity curve is characterized by a fast, seconds-long, rise and an equally fast decline, with a minute- or hour-long plateau in between. For a near-Earth star (distance $D\lesssim 1$ kpc) the echo will be observable at near future gamma ray telescopes with an effective area of $10^{3}~\mathrm{cm^2}$ or larger.  Its observation will inform us on the envelope size and composition. In conjunction with the direct detection of the neutrino burst, it will also give information on the neutrino emission away from the line of sight and will enable tests of neutrino  propagation effects between the stellar surface and Earth.
\end{abstract}
\section{Introduction}
A core collapse \sn\ is the most powerful \n\ emitter known so far. The $\sim 10$ s-long burst of thermal \ns\ emitted from the outskirts of the collapsed core is the main cooling mechanism, and is a powerful diagnostic tool of the physics that takes place in the very dense and hot region deep inside the star. 

Interestingly, one of the best \sn\ \n\ detectors is the most abundant element in the universe, Hydrogen. Indeed, the process of inverse $\beta$ decay (IBD), $\barnue +\ p \rightarrow\ \positron +\ n $, has a relatively large, well known, cross section~\citep{Vogel:1999zy,Strumia:2003zx} and, depending on the type of detector, it can provide information on the energies and arrival times of the individual \ns\ detected. This simple, reliable method has found application in water and liquid scintillator detectors \citep{Scholberg:2012id}, and it was used in the first and only detection of \sn\ \ns, the burst from SN1987A \citep{Kamiokande-II:1987idp,Bionta:1987qt,Alekseev:1988gp}. Its evolution has been driven by the need of having larger detector masses; e.g., about ${\mathcal O}(100)$ kt mass of water is needed for high statistics detection of supernovae beyond our galaxy. 

\begin{figure}
\begin{center}
{\includegraphics[width=0.46\textwidth,angle=0]{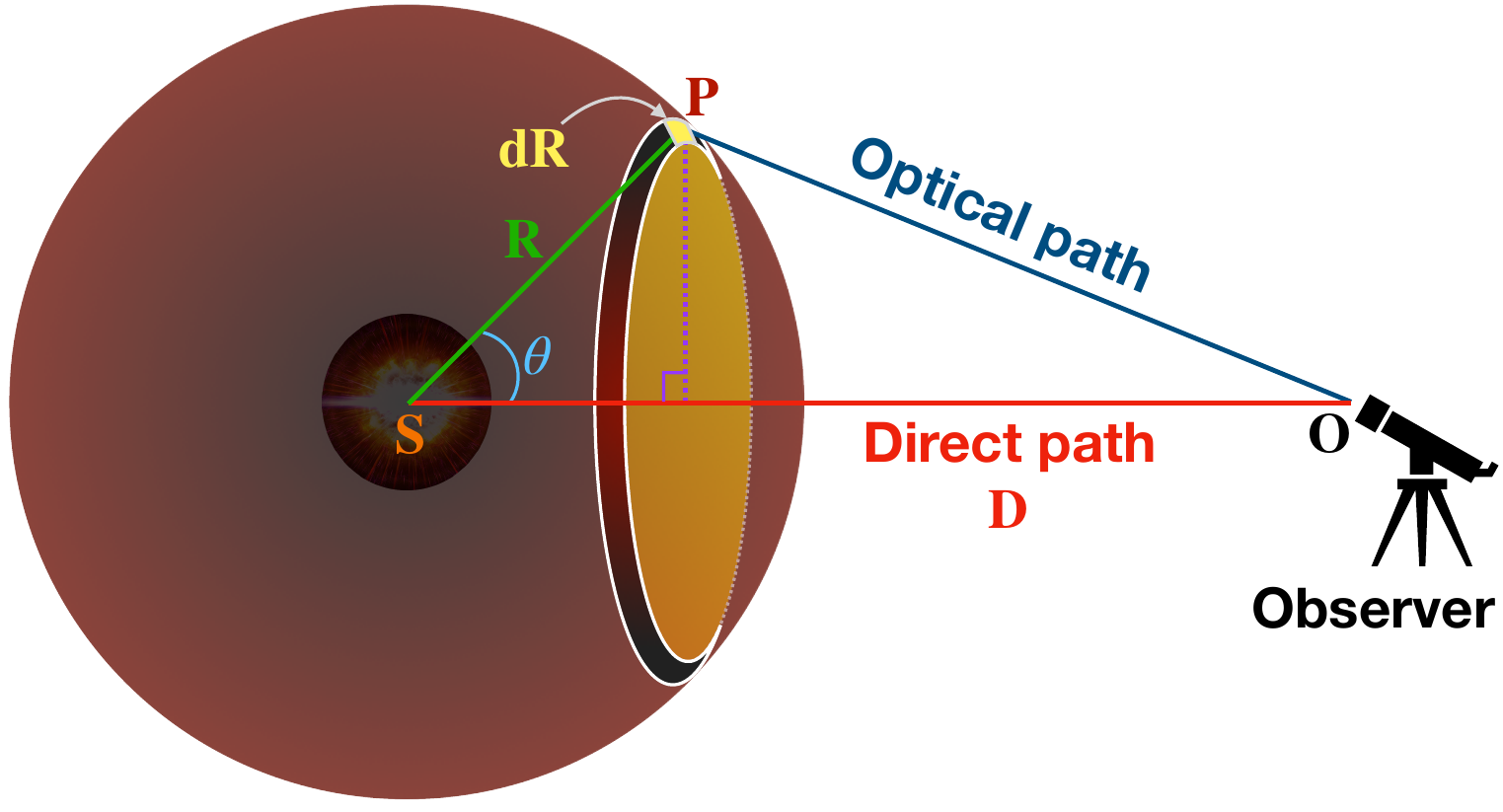}}
\caption{\label{fig:cartoon} 
Geometry of a gamma ray echo.}
\end{center}
\end{figure}

The concept of Hydrogen as detector leads to an idea: why not use the vast mass of Hydrogen \emph{in or near the star itself} 
as detector?  
This question was first studied several decades ago, when it was observed in \cite{1975Ap&SS..35...23B,1999NCimC..22..115R} that inverse beta decay in the hydrogen envelope of a collapsing star leads to a transient signal of positron annihilation  ($e^+ + e^- \rightarrow \gamma + \gamma $) signatures, mainly in the form of 0.511 MeV gamma rays \citep{Lu:2007wp}\footnote{Other channels were considered, but the most dominant signal was the $0.511$ MeV $\gamma-$rays.}.
Due to the geometry of the system (Figure~\ref{fig:cartoon}), these gamma rays arrive at Earth as an \emph{echo}, spread over a characteristic time $\Delta t \sim R/c$ (with $R$ being the star's radius), relative to the \n\ burst. In those early studies, the predicted luminosity of this echo was considered too low for observation, and therefore this phenomenon was largely ignored since.

In this letter, we present a modern study of the gamma ray echo of a \sn\ \n\ burst. 
There are two main elements of novelty. The first is a prediction of the gamma ray light curve, and its dependence on the main parameters.  The second is the discussion of the potential of upcoming gamma-ray surveys to observe the echo from nearby core-collapse supernovae, and extract important information from it.  With improved, next-generation, gamma ray telescopes like COSI~\citep{Tomsick:2021wed} (already funded), AMEGO~\citep{AMEGO:2019gny,Kierans:2020otl}, and AMEGO-X~\citep{Caputo:2022xpx}, detecting this 511 keV signal 
will soon be a realistic possibility.

The paper is organized as follows. We discuss the formalism of the gamma ray echo, along with its time-dependent flux in section~\ref{sec:formalism}. The detectability of the echo, and relevant backgrounds are discussed in section~\ref{sec:detectability}. We summarize and discuss future prospects in section~\ref{sec:disc}.

\section{Formalism}
\label{sec:formalism}

To fix the ideas, we focus on the case which is most favorable for detection, where the gamma rays originate near the surface of the star, and propagate without absorption to Earth (attenuation will be discussed below).  
Let's begin by estimating the total photon flux at Earth.
We assume a spherically symmetric star, and model the flux of $\barnue$ reaching its surface (after flavor conversion, see, e.g. \citealt{Duan:2009cd} for a review) as having total energy $E_{\nu,tot}=5~10^{52}$ ergs. 

The commonly used ``alpha spectrum" is assumed for its energy distribution \citep{Keil:2002in,Tamborra:2012ac}, with the first two momenta being $\langle E_\nu \rangle = 15$ MeV and $\langle E^2_\nu \rangle = 293.2 ~\mathrm{MeV^2}$  (corresponding to the shape parameter $\alpha =2.3 $, where $(1+\alpha)^{-1}=\langle E^2_\nu \rangle/\langle E_\nu \rangle^2 -1$). For simplicity, we use a time-independent spectrum, so the total number of $\barnue$ emitted is, simply, $N_\nu=E_{\nu,tot}/\langle E_\nu \rangle\simeq 2.1~10^{57}$.    
Two cross sections are relevant here: one is the spectrum-averaged IBD cross section, $\langle \sigma_{\rm IBD} \rangle=2.05~10^{-41}~\mathrm{cm^2}$, the other is for the Compton scattering of gamma rays, $\sigma_{\rm C}=3~10^{-25}~\mathrm{cm^2}$~\citep{rybickilightman}, which is the main channel of photon absorption at the energies of interest \citep{Lu:2007wp}.  As an approximation, we consider that the emerging flux of gamma rays is entirely due to the $\barnue$ that interact in the outermost layer of the star, a very thin shell of width equal to the gamma ray Compton optical depth ($l_{\rm C}\sim {\mathcal O}(10^9)$ cm, $l_{\rm C}\ll R$, see \citealt{Lu:2007wp}). One can then express the number of positrons produced in this layer, and the corresponding number of gamma rays from positron propagation that leave the star as \citep{1999NCimC..22..115R}: 
\begin{eqnarray}
    &&N_{+} \simeq \frac{Y_p \langle \sigma_{\rm IBD} \rangle }{Y_e \sigma_{\rm C}} N_\nu \simeq 1.25~10^{41} ~; \nonumber \\
     && N_{\gamma} =\frac{1}{2} \eta_\gamma N_{+}\sim N_{+}~,
    \label{eq:num}
\end{eqnarray}
where $Y_p\sim 1$ and $Y_e\sim 1$ are the proton and electron fractions in the stellar matter; $\eta_\gamma\sim 2$ is the effective  number of 0.511 MeV gamma rays produced per positron (see below), and the factor $1/2$ accounts for the fact that half of the gamma rays propagates inwards and is absorbed. 
By symmetry arguments, it follows immediately that the time-integrated photon flux at Earth is~\footnote{Our result is a factor of 4 larger than the one in \cite{Lu:2007wp}. We were unable to trace the origin of the difference; still, the geometric factor $4\pi D^2$ in Equation~(\ref{eq:totalflux}) is substantiated by the extended derivation in \cite{FAVORITE201629}.}
\begin{equation}
\Phi_{\gamma, tot}=\frac{N_{\gamma}}{4 \pi D^2}\sim 10^{-3}~\mathrm{cm^{-2}}  \left(\frac{D}{\mathrm{ kpc}} \right)^{-2}~.
\label{eq:totalflux}
\end{equation}
We estimate the uncertainty on $N_{+}$ and $N_\gamma$ to be about $\sim 50\%$ in either direction due to the uncertainty in the neutrino spectrum parameters. 

\subsection{Time-dependent gamma ray flux}
To compute the expected time-dependent gamma ray flux and its energy spectrum, it is necessary to consider a specific stellar environment and model the positron propagation in detail. Here we follow the extensive discussion in \cite{Lu:2007wp}, where a precise estimate for $\eta_\gamma$ is found.  
There, the values $Y_p=0.7$ and $Y_e=0.85$ are used. It is shown that  positron annihilation is the main channel of gamma ray production, dominating over the secondary channel -- the emission of $2.22$ MeV gamma rays from neutron capture -- by roughly two orders of magnitude in flux. 
For a hydrogen envelope in thermodynamic equilibrium at temperature $T \sim 10^4$ K and density $\rho \sim 10^{-8}\ \rm g~ cm^{-3}$, it was found 
that positron thermalization is -- in the vast majority of cases -- fast, occurring over a typical time scale of $\sim 10^{-2}$ s,  
with the excitation of free electrons being the dominant energy loss mechanism. Direct positron annihilation with free electrons is the main absorption process, and the probability that it occurs before thermalization is estimated to be $P_{fa}\simeq 0.1$ (see eq. (53) in \cite{Lu:2007wp}). Therefore, roughly a fraction  $P\sim 1 - P_{fa}\simeq 0.9$ of all positrons annihilate after thermalization. A more detailed calculation, which includes several other energy loss and absorption processes, and the formation of positronium states, leads to $P\simeq 0.87$  (see fig. 3 and related text in \cite{Lu:2007wp}). 
The annihilation of thermalized positrons results
in a gamma ray spectrum that is centered at $E_\gamma =0.511$ MeV, with width (full width at half maximum) $\Delta E_\gamma \simeq 2$ keV.
Accounting for the fact that each annihilation produces two photons, we therefore estimate $\eta_\gamma =2 P =1.74$ which will be used here as reference value.

Let us now describe the expected gamma ray lightcurve at Earth, for a star at distance $D$, and a given neutrino number luminosity  $L_\nu(t)=d N_\nu/dt$. 
The gamma ray flux at Earth, $\Phi_\gamma$, is obtained by integrating over the visible surface of the star, and by considering that photons reach the detector with a time delay that increases with their angular distance, $\theta$, from the line of sight (see Figure~\ref{fig:cartoon}). We find the expression:
\begin{widetext}
\ba
\Phi_\gamma(t,R,D) &=&  \frac{\eta_\gamma}{8 \pi D^2}\  \frac{Y_p \langle \sigma_{\rm IBD} \rangle }{Y_e \sigma_{\rm C}} \int_0^1 L_\nu \Big(t - \frac{R}{c} (1-\cos{\theta})\Big) d(\cos{\theta})~, \nonumber \\
&=& \frac{\eta_\gamma}{8 \pi D^2}\  \frac{Y_p \langle \sigma_{\rm IBD} \rangle }{Y_e \sigma_{\rm C}} \int^\infty_{-\infty} B(y)L_\nu(t-y)dy ~,
\label{eq:gflux}
\ea
\end{widetext}
which is consistent with Equation~(\ref{eq:num}), and where $c$ is the speed of light, and $B$ is a  box function normalized to 1: $B(y)=c/R$ for $0\leq y \leq R/c$, and $B(y)=0$ elsewhere. The second line of Equation~(\ref{eq:gflux}) emphasizes that the echo is described by a convolution operation (see, e.g., the formalism in \citealt{Dwek:2021nsk}). 
Here $t=0$ is set to be the start of the \n\ burst as observed at Earth. 

\begin{figure*}[thb]
\begin{center}
{\includegraphics[width=0.9\textwidth]{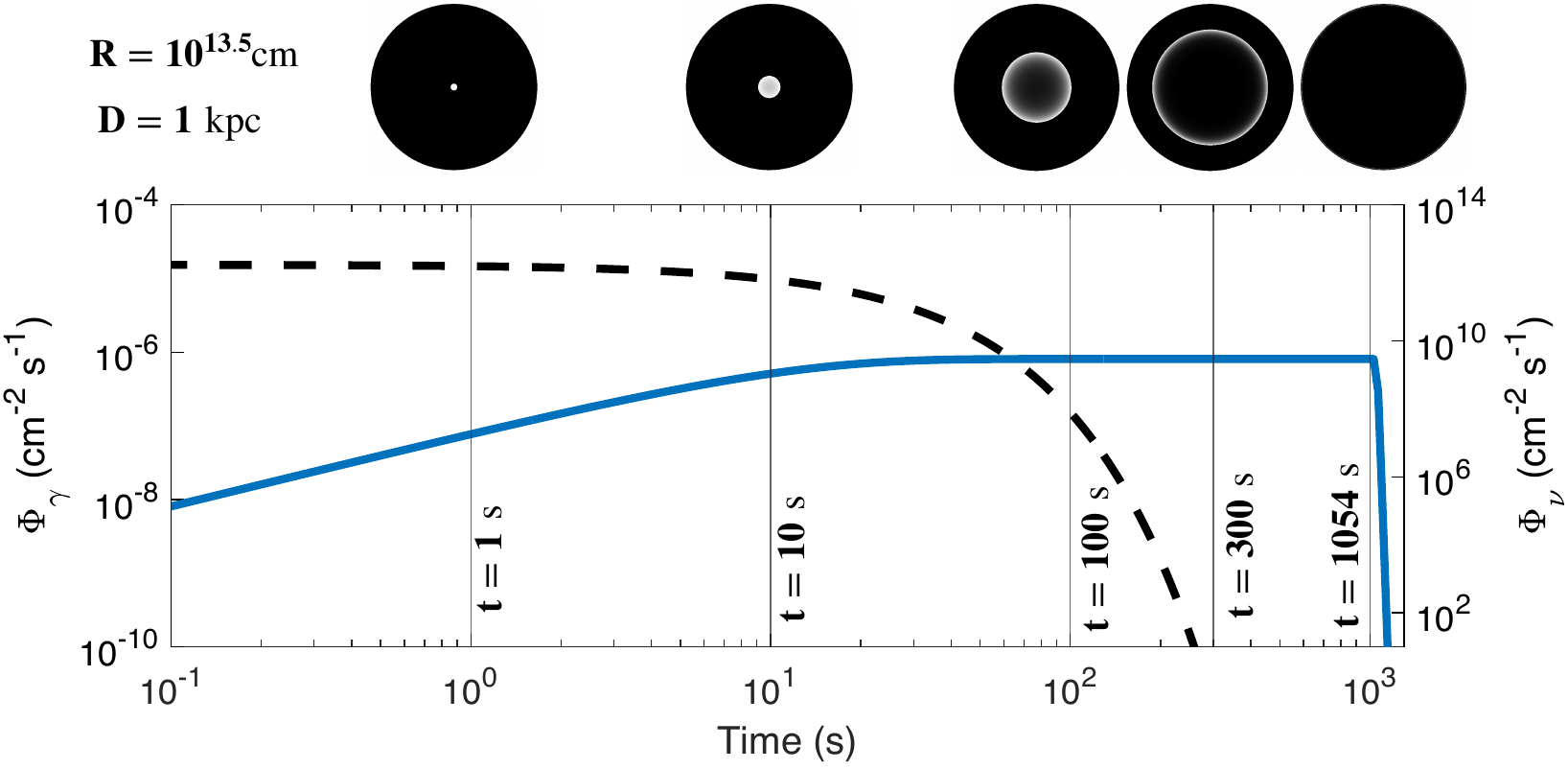}}
\caption{\label{fig:snapshots} 
Example of neutrino (dashed, vertical axis on the right) and gamma ray  (solid, vertical axis on the left) lightcurves. See legend for the stellar radius and distance to Earth. Included are time snapshots of the star, showing the part that shines in gamma rays  as seen by an observer at Earth (regions in white in the black disks). }
\end{center}
\end{figure*}
\subsection{Results}
Using the expression in Equation~\ref{eq:gflux}, the gamma ray flux can be computed for specific scenarios of neutrino emission. 
For illustration, we model the \n\ luminosity as a truncated  exponential: 
\begin{equation}
    L_\nu(t) = \begin{cases}
    L_0 e^{-t/\tau} &  0\leq t \leq t_0 \\
    0 & \textrm{ elsewhere }
    \end{cases}~,
    \label{eq:lumfun}
\end{equation}
where the limit $t_0 \rightarrow +\infty$ (no truncation) well approximates the case of a neutron-star forming collapse, where the proto-neutron star cools smoothly by \n\ emission over several tens of seconds. The case $t_0 \lesssim 1$ s could describe a collapse with direct black-hole formation (failed supernova, see, e.g. \citealt{OConnor:2010moj,Pejcha:2014wda,Ertl:2015rga}), for which the \n\ emission is truncated sharply when the neutrinosphere falls within the gravitational radius. Here we take $t_0=1$ s (for black hole formation) and $\tau=3$ s.
The description in Equation~(\ref{eq:lumfun}) captures the main features of the luminosity curve over a multi-second timescale, which is sufficient for the present scope. We note that fast  fluctuations of $L_\nu(t)$ (over a time scale 0.1 s or less) such as those expected in the first second or so of the \n\ burst \citep{Foglizzo:2001ke,Blondin:2002sm,Foglizzo:2002hi} would in any case be smoothed out by the integration in Equation~(\ref{eq:gflux}), and therefore have a negligible effect on the gamma ray lightcurve.

\begin{figure*}[thb]
\begin{center}
{\includegraphics[width=0.7\textwidth]{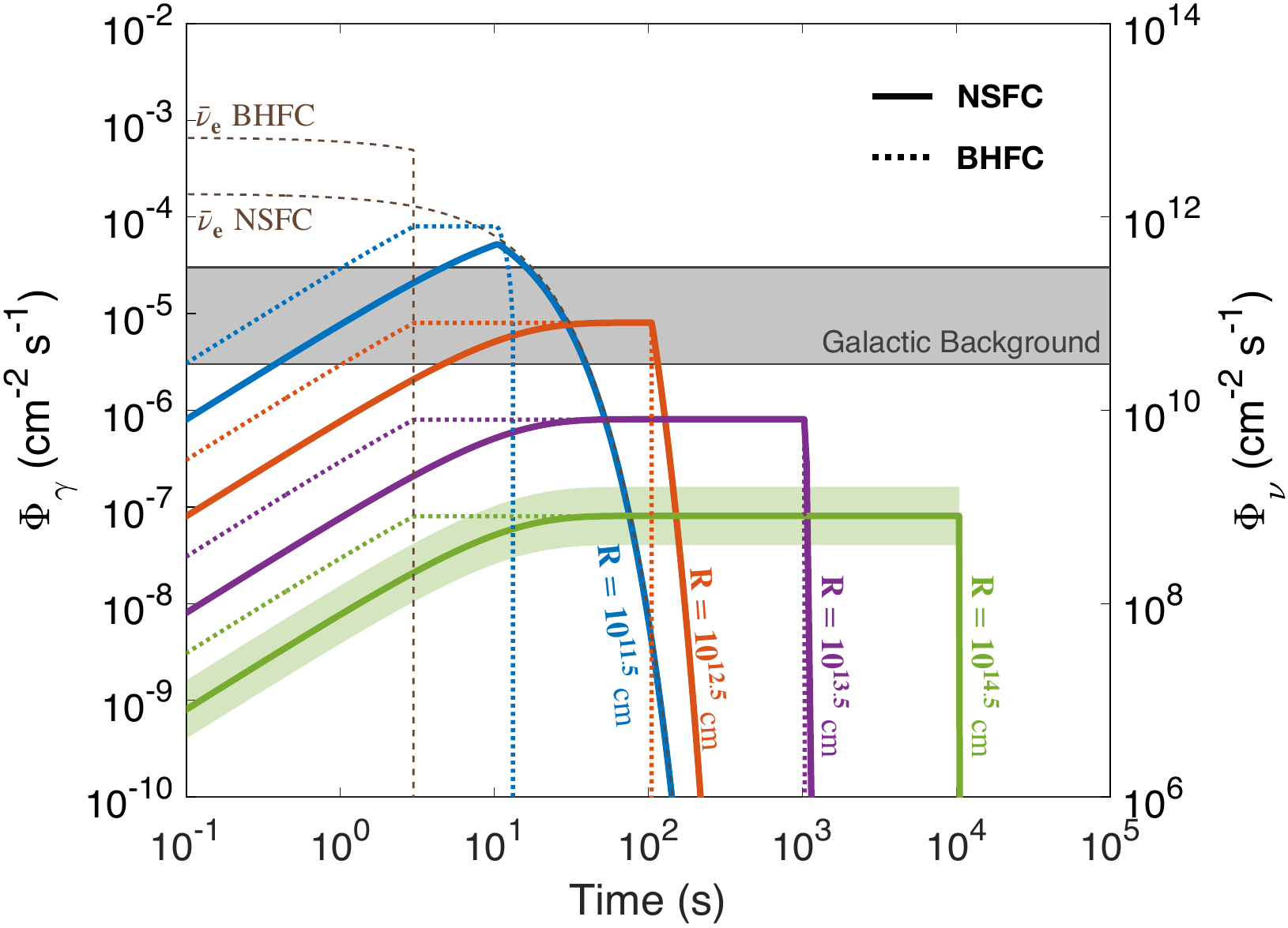}}
\caption{\label{fig:lightcurve} 
  Predicted gamma ray lightcurves for  
  a neutron-star-forming collapse (NSFC, solid lines) and a direct black-hole-forming collapse (BHFC, dotted lines), at distance $D=1$ kpc and different stellar radii (labels on curves). 
  Dashed: $\barnue$ flux for the two cases (see right vertical axis for scale). 
  For all the results, we used $E_{\nu,tot}=5~10^{52}$ ergs for the total energy emitted in $\barnue$. The galactic background is shown as well. For both signal (one case only, for illustration) and background, the shadings represent the uncertainties discussed in the text. 
  }
\end{center}
\end{figure*}
The result for the case of a neutron-star forming collapse is 
sufficiently simple, and is given by: 
\begin{align}
\Phi_\gamma(t,R,D) &= \Phi_0(R,D) \begin{cases}
    0 &  t \leq 0 \\
    \left(1-e^{-\frac{t}{\tau}}\right) & 0 < t\leq R/c \\
   \left(e^{\frac{R}{c \tau}}-1\right)e^{-\frac{t}{\tau}} &  t > R/c
    \end{cases}~,
    \label{eq:phioft}\\
\Phi_0 (R,D) &= \frac{\eta_\gamma}{8\pi D^2}\frac{Y_p \langle \sigma_{\rm IBD} \rangle }{Y_e \sigma_{\rm C}} \frac{L_0 c\tau}{R} \nonumber \,.
\end{align}
The results are shown in Figure~\ref{fig:snapshots} and \ref{fig:lightcurve} (solid curves). 
The expression in Equation~(\ref{eq:phioft}) describes the ``phases" of the star as seen by an observer at Earth (see illustration in Figure~\ref{fig:snapshots}): first, there is an increase in flux ($0 \leq t\leq R/c$), when the surface of the star becomes bright in a circle around the line of sight, and the circle expands. After a time comparable with the \n\ emission timescale, $t \sim 3 \tau$, the luminosity of the echo has reached a plateau. This behavior describes the phase where the gamma-ray emitting region of the star -- as seen at Earth --  is made of an expanding annulus where the intensity of emission is at its maximum, whereas the region near the line of sight emits less intensely due to the decline on $L_\nu$. 
The plateau lasts until $t= R/c$, when the entire visible face of the star has become bright in gamma rays; at later times the star still appears completely illuminated, but the gamma ray flux declines over a timescale $\sim \tau$ because all the points on its surface are receiving a \n\ flux that is past its peak luminosity. 

As shown in Equations (\ref{eq:gflux}) and (\ref{eq:phioft}), and  Figure~\ref{fig:lightcurve}, the echo becomes fainter and longer for larger envelope radii; for a reference radius  $R=10^{13.5}$ cm and distance $D=1$ kpc to the star, we estimate a duration of $R/c \simeq 10^3$ s (approximately 17 minutes) and maximum flux $\Phi_\gamma\sim 10^{-6}~\mathrm{ cm^{-2} s^{-1}}$. 

\begin{figure}
\includegraphics[width=0.45\textwidth]{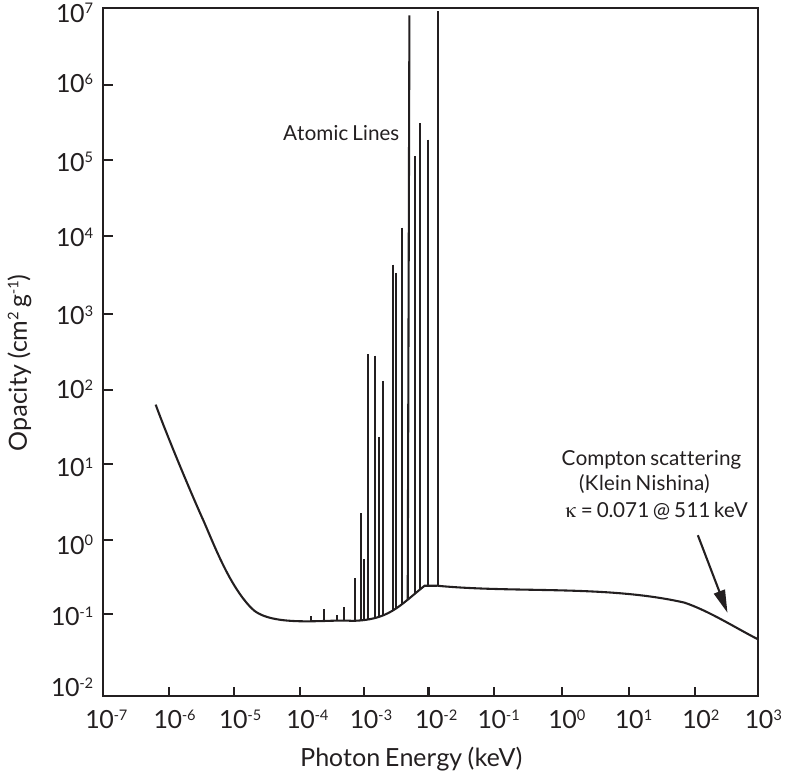}
\caption{\label{fig:opacity_fig} 
A schematic graph illustrating the opacities of the ISM corresponding to the various values of photon energy. The regime of interest for this work ($0.511$ MeV) is shown with an arrow. 
}
\end{figure}
The case of a failed \sn\  -- for which the analytical result is complicated, and will be omitted for simplicity --   is described in Figure~\ref{fig:lightcurve} (dotted lines). Qualitatively, the behavior is similar to the previous case, with the difference that the transition between phases is sharper, reflecting the sudden drop of $L_\nu$.  In this case, an observer at Earth would see a sharp boundary between a fully illuminated annulus and a completely dark circle centered at the line of sight.  Note that, by construction, the total energy emitted in \ns\ is the same for the two types of collapses. Therefore, due to the shorter time scale of the emission, for a failed \sn\ the rise phase of the echo is more luminous and could be more easily observed.



\begin{figure*}[thb]
\begin{center}
{\includegraphics[width=0.75\textwidth]{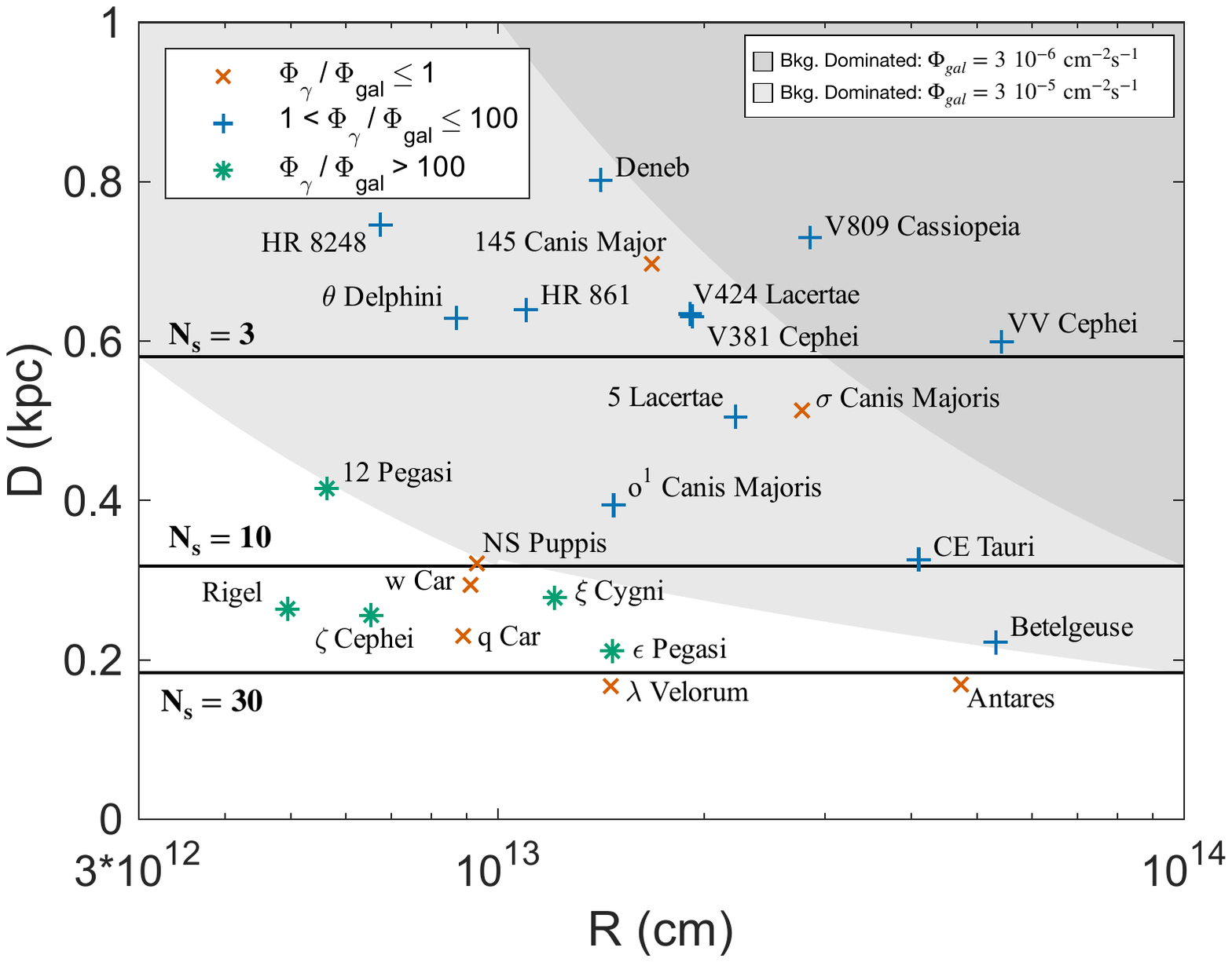}}
\caption{\label{fig:detection} The detectability of the echo depending on the distance, $D$, and the stellar radius, $R$, for a telescope with $A=10^3~\mathrm{cm^2} $ area. The horizontal (solid) lines correspond to a fixed number of signal events $N_s$ (numbers on curves). The  region outside the shaded areas is where the signal can be distinguished from the background (see text), assuming a (fixed) galactic background flux $\Phi_{gal}=3~10^{-6}~\mathrm{ cm^{-2} s^{-1}}$ (darker shading)  
and $\Phi_{gal}=3~10^{-5}~\mathrm{ cm^{-2} s^{-1}}$  (lighter shading) respectively. The markers represent the nearby stars for which the radii are known, and correspond to different intervals of realistic (direction-specific) signal-to-background ratio; see legend and Table \ref{tab:presncandidates}. 
}
\end{center}
\end{figure*}
\subsection{Attenuation}
Let us briefly discuss the attenuation of the emitted gamma rays due to propagation in the 
circumstellar medium (CSM) and interstellar medium (ISM). The attenuation factor is $\eta_{abs}=\exp \big( - \int ds\ \kappa_\nu \rho\big)$, where $\kappa_{\nu}$, and $\rho$ are the monochromatic opacity and density of the medium respectively, and the integral is performed along the line of sight. For simplicity, we approximate the CSM using the radially averaged density and temperature profiles from~\cite{2013A&A...559A..69G}, where typical values are in the range: $-29\lesssim$~log$(\rho/\textrm{g/cm$^{3}$})\lesssim -22.5$, $2.5\lesssim$~log$(T/\textrm{K})\lesssim 8.5$. For these, the $0.511$ MeV photon interactions are dominated by Compton scattering (in the Klein-Nishina regime): $\kappa_{\nu} \approx 0.071$ cm$^{2}$/g. Absorption by atoms and ions is negligible, being most effective at lower energy (see Figure~\ref{fig:opacity_fig}, where the corresponding spectral lines are shown). 
We find that  attenuation is practically negligible; for example $\eta_{abs}-1\simeq 1.5~ 10^{-6}$ for a CSM of $15$ pc width, and  $\eta_{abs}-1 \simeq 2.2~10^{-4}$ for propagation in the ISM over a distance of $1$ kpc.  
\section{Detectability}
\label{sec:detectability}
In this section, we focus on the detectability of the gamma ray echo and discuss the relevant backgrounds. 
The echo is detectable \emph{in principle} if it produces at least one photon signal in a detector at Earth: $N_{s} = A \Phi_{\gamma, tot}\gtrsim 1 $, where $A$ is the effective area of the telescope, which we assume to be pointing in the direction of the star (incidence angle $\theta_i=0$). Using Equation~(\ref{eq:totalflux}), we obtain the detectability condition (for fixed neutrino flux parameters): 
\begin{equation}
  \left( \frac{A}{\mathrm{10^3~ cm^2}} \right)  \left(\frac{D}{\mathrm{ kpc}} \right)^{-2}   \gtrsim 1~.
   \label{eq:cond1}
\end{equation}

Current or upcoming instrument typically have $A\lesssim 10^2~\mathrm{cm^2}$ (see, e.g, COSI \citealt{Tomsick:2021wed} and 511-CAM \citealt{Shirazi:2022taf}), and would therefore only be able to observe an echo from a very nearby collapse with an exceptionally luminous neutrino emission. In particular, the NASA funded detector COSI has  $A>20\ \rm cm^2$ as design sensitivity at $511$ keV, along with a $25$\% field of view of the sky and an angular resolution of $< 4.1^\circ$. 
With a factor of 2 improvement on its design specification ($A=40\ \rm cm^2$), COSI would be able to see the echo from stars at $D\sim 0.2$ kpc, like supernova candidates Betelgeuse 
and $\epsilon$ Pegasi.
For the largest telescopes of the next generation, AMEGO \citep{AMEGO:2019gny,Kierans:2020otl} ($A \sim 3~10^3$ cm), and GECCO ($A \sim 800$ cm) \citep{Orlando:2021get}, 
the echo might be visible for stars within a radius of 1 kpc or so, where about 31 supernova candidates are located \citep{Mukhopadhyay:2020ubs}.
\subsection{Backgrounds}
Realistically, detection requires the signal to be distinguishable above the relevant backgrounds, mainly the diffuse galactic background at $E_\gamma = 0.511$ MeV, which peaks in the direction of the Galactic Center (see, e.g., \citealt{2013AstRv...8c..19D,2016JPhCS.703a2001R,2021PASA...38...62D,2021ExA....51.1175F}). We do not consider possible contamination of an echo signal from other nearby sources like gamma ray bursts, other supernovae, supernova  remnants, low mass X-ray binaries, and others. Positron annihilation from competing processes in the stellar envelope is unlikely to contribute, as the temperature of the envelope is well below the threshold for thermally producing electron-positron pairs, and  positrons from (post-collapse) radionuclides near the core of the star would be located too deep to affect the echo over its short timescale.

Let us estimate the diffuse galactic background at $E_\gamma = 0.511$ MeV. We take the value $d\Phi_{gal}/d\Omega \simeq 3~ 10^{-4}~\mathrm{ cm^{-2}~s^{-1}~sr^{-1}}$ as a reference for this flux away from the galactic center (see results in \citealt{Skinner:20156m}, for galactic latitude and longitude $b=0^\circ$ and $|l|=60^\circ$).
Taking the published angular resolution of AMEGO, $\delta \theta\simeq 3^\circ$~\footnote{Available at \href{https://asd.gsfc.nasa.gov/amego/technical.html}{AMEGO technical sheet}.} (corresponding to a solid angle  $\delta \Omega =  \pi (\delta \theta)^2 \simeq 10^{-2}$ sr), we obtain a flux $\Phi_{gal} \simeq 3~10^{-6}\mathrm{ cm^{-2}~s^{-1}}$, which is comparable to the maximum value predicted for the echo for $R\gtrsim 10^{13}$ cm and $D=1$ kpc (see Figure~\ref{fig:lightcurve}).  This indicates that realistically, a distance below the kpc scale is needed for a robust identification of the echo. 
A rough criterion for the signal to be detectable above the background is that the number of signal photon counts exceeds the one of the background over the duration of the echo: $N_s/N_B  \gtrsim 1$, where $N_B\simeq \Phi_{gal} A R/c$ \footnote{
For simplicity, here we refer to the case where the signal flux is dominated by the plateau phase, i.e., $R/c \gg  10$ s, which is realized for the nearby stars considered here (Table \ref{tab:presncandidates}). In this case
the condition $N_s/N_B  \gtrsim 1$ is well approximated by $\Phi_\gamma/\Phi_{gal} \gtrsim 1$, with $\Phi_\gamma$ being the flux at the plateau. }. Numerically, we get:
\begin{equation}
 \left( \frac{\Phi_{gal}}{3~10^{-6}\mathrm{ cm^{-2}~s^{-1}}}\right)  \left( \frac{D}{\mathrm{1 kpc}}\right)^2  \left(\frac{R}{10^{13}\mathrm{cm }} \right) \lesssim 0.74 ~.
   \label{eq:cond2}
\end{equation}
If $N_s \gtrsim 10$,  one can use the (less stringent) requirement that the signal exceeds a 3$\sigma$ Gaussian fluctuation of the background:  $N_s> 3 \sqrt{N_B} $. 
\subsection{Prospects for detection amongst nearby supernova candidates}
Since a detection is possible only for near-Earth collapses, we have examined the known hydrogen-rich \sn\ candidate stars within a radius of 1 kpc; they are listed in Table~\ref{tab:presncandidates} (taken from \citealt{Mukhopadhyay:2020ubs} with some updates, see table caption). For each, we report the estimated distance and radius, the value of $\Phi_{gal}$ in the direction of the star and our result for $N_s/N_B$. 
The direction-specific $\Phi_{gal}$ is evaluated from~\cite{Skinner:20156m} (see Figure~2 there).
In the absence of a a detailed three dimensional model of the galactic background,  
the angular separation between the galactic center and the star has been calculated only along the galactic longitude ($l$), assuming latitude $b = 0$. 
For stars with angular separation exceeding $ 90^\circ$, we take the fixed value $\Phi_{gal}=3.0\ 10^{-7} \rm cm^{-2} s^{-1}$, which is an overestimate, and therefore leads to conservative conclusions. 
 
The conditions for detectability, Equations (\ref{eq:cond1}) and (\ref{eq:cond2}) (and its extension to Gaussian statistics, see above), are  illustrated in Figure~\ref{fig:detection}, for two fixed values of the galactic background flux, $\Phi_{gal} = 3 \ 10^{-6} \rm cm^{-2}s^{-1}$ and $\Phi_{gal} = 3 \ 10^{-5} \rm cm^{-2}s^{-1}$. 
The candidate stars in Table \ref{tab:presncandidates} are shown as well, color coded according to the realistic (direction-specific) signal-to-background ratio given in the Table. From the figure, it appears that the detection of the echo is possible - with a telescope having $A \gtrsim 10^3~\mathrm{cm}$  - for several candidates. Many of them are in a fortunate location, where the galactic background is low. Examples are Betelgeuse and Rigel, for which the angular separation from the galactic center is $\sim 157^\circ$ and $\sim 123^\circ$ respectively, and therefore the background is overestimated. Rigel is also favored by is relatively compact size, $R\simeq 5~10^{12}$ cm, which implies a higher peak of the echo flux (see Figure~\ref{fig:lightcurve}). In contrast, Antares - which is similar in distance and radius to Betelgeuse - is disfavored by its proximity to the Galactic Center, which causes a higher background ($\Phi_{gal} \simeq 2.3\ 10^{-5} \rm cm^{-2}s^{-1}$) and lower signal-to-background ratio.

\section{Discussion and future prospects}
\label{sec:disc}
%
The observation of the gamma ray echo could reveal information that would otherwise be inaccessible. In particular, the echo provides information on the $\barnue$ flux passing through the stellar surface facing the observer. 
Therefore, the comparison with the detected \n\ burst would test the intensity of the \n\ emission \emph{away from the line of sight}. It would also  allow to search for exotic effects that might affect the \ns\ between the star and Earth, like \n\ decay, conversion into sterile states, scattering on Dark Matter, etc. 
Our predicted gamma ray flux might serve as a reference for searches of gamma-ray-producing effects beyond the Standard Model, like axion-photon conversion (see, e.g. \citealt{Chattopadhyay:2023nuq} for a related idea). 
The echo could also provide independent estimates of the progenitor star's features, mainly the radius and envelope composition. 
Its initial, rising phase could also contribute to early alerts of the collapse, preceding the explosion (or collapse into black hole) of the star. 

This work can be extended in many ways, to obtain more realistic predictions. For example, post-main sequence massive stars undergo radial pulsations with periods of $100-1000$ days \citep{goldberg_2020_aa}, which would change the envelope structure, and therefore influence the lightcurve of the gamma ray echo. One could also use numerical results for the stellar envelope structure and composition and for the \n\ luminosity, that might include deviations from spherical symmetry. 
Different, more realistic forms of the time profile of the \n\ luminosity could be examined (e.g., a power-law form, see \citealt{Suwa:2020nee}).
Including secondary branches of positron propagation, and the contribution of the layers of the star deeper than the gamma ray optical depth may result in more optimistic predictions for the gamma ray flux. The idea of a gamma ray echo could be extended to other set-ups, for example involving supernova progenitors with Carbon- or Oxygen-rich envelopes - where gamma rays can be produced by neutrino-nucleus scattering - , or envelope-stripped stars where the echo might be due to a detached hydrogen shell (for example, the shell between two companion stars, see, e.g., \citealt{Pejcha:2021qfk}). The latter, however, might give very faint echoes due to their more extended structure (larger $R$).

In conclusion, we have presented a modern rendering of the idea of supernova neutrinos producing a gamma ray echo. This phenomenon is conceptually interesting, because the neutrinos take the unusual role of being the source of an electromagnetic signal, and is also attractive as a realistic target of observation for future large gamma ray telescopes with sub-MeV capability. It adds another facet to the very rich landscape of multi-messenger astronomy.

\begin{deluxetable*}{cccccccc}
\tablecolumns{8}
\tablecaption {Nearby red and blue supergiants and their estimated radii, distances and positions; adapted from \cite{Mukhopadhyay:2020ubs}.  Also given are  the angular separation from the galactic center (along $l$, assuming $b = 0$), the associated $511$ keV galactic background, and the predicted signal-to-background ratio.   We only list stars to which our scenario applies, namely, stars that have a hydrogen envelope and for which the radius $R$ is known. } 
\label{tab:presncandidates} 
\tablehead{
\colhead{Common Name} & \colhead{$D$ (in kpc)} & \colhead{$R$ ($\times 10^{13}\rm cm$)} & \colhead{RA} & \colhead{Dec} & \colhead{$l$ Sep.} & \colhead{$\Phi_{gal}$ ($\rm cm^{-2} \rm s^{-1}$)} & \colhead{$N_s/N_B$} 
}
    \startdata
     $\lambda$ Velorum & $0.167 \pm 0.003$   & $1.46$~\tablenotemark{a}  & 09:07:59.76 & -43:25:57.3 & $2.61^\circ$ & $8.64 \times 10^{-5}$ & 0.732 \\
     Antares/$\alpha$ Scorpii & $0.169 \pm 0.030$  & $4.73$~\tablenotemark{b}  & 16:29:24.46 & -26:25:55.2 & $14.94^\circ$ & $1.97 \times 10^{-5}$ & 0.968 \\
     $\epsilon$ Pegasi & $0.211 \pm 0.006$   & $1.47$~\tablenotemark{c} & 21:44:11.16 & +09:52:30.0 & $148.76^\circ$ & $2.58 \times 10^{-7}$ & $1.53 \times 10^{2}$ \\
     Betelgeuse &  $0.222 \pm 0.040$  & $5.32$~\tablenotemark{d} & 05:55:10.31 & +07:24:25.4 & $157.43^\circ$ & $2.58 \times 10^{-7}$ & 38.1 \\
     q Car/V337 Car &  $0.230 \pm 0.020$   & $0.890$~\tablenotemark{e} & 10:17:04.98 & -61:19:56.3 & $2.95^\circ$ & $8.30 \times 10^{-5}$ & 0.659 \\
     $\zeta$ Cephei &  $0.256 \pm 0.006$   & $0.654$~\tablenotemark{f}  & 22:10:51.28 & +58:12:04.5 & $178.74^\circ$ & $2.58 \times 10^{-7}$ & $2.33 \times 10^{2}$\\
     Rigel/$\beta$ Orion &  $0.264 \pm 0.024$   & $0.494$~\tablenotemark{g}   & 05:14:32.27 & -08:12:05.90 & $122.61^\circ$  & $2.58 \times 10^{-7}$ & $2.90 \times 10^{2}$ \\
      $\xi$ Cygni &  $0.278 \pm 0.029$   & $1.21$~\tablenotemark{f}  & 21:04:55.86 & +43:55:40.3 & $178.05^\circ$  & $2.58 \times 10^{-7}$  & $1.07 \times 10^{2}$ \\
     w Car/V520 Car & $0.294 \pm 0.023$   & $0.913$~\tablenotemark{h}  & 10:43:32.29 & -60:33:59.8 & $1.09^\circ$ & $1.11 \times 10^{-4}$ & 0.296 \\
     NS Puppis &  $0.321 \pm 0.032$  & $0.932$~\tablenotemark{f}  & 08:11:21.49 & -39:37:06.8 & $3.50^\circ$ & $7.84 \times 10^{-5}$ & 0.342 \\
     CE Tauri/119 Tauri & $0.326 \pm 0.070$   & $4.10$~\tablenotemark{i} & 05:32:12.75 & +18:35:39.2 & $167.71^\circ$ & $2.58 \times 10^{-7}$ & 22.9 \\
     $o^1$ Canis Majoris & $0.394 \pm 0.052$   & $1.47$~\tablenotemark{f}  & 06:54:07.95 & -24:11:03.2 & $22.79^\circ$ & $5.49 \times 10^{-6}$ & 2.05 \\
     12 Pegasi & $0.415 \pm 0.031$   & $0.564$~\tablenotemark{f}     & 21:46:04.36 & +22:56:56.0 & $159.50^\circ$ & $2.58 \times 10^{-7}$ &  $1.03 \times 10^{2}$  \\
     5 Lacertae & $0.505 \pm 0.046$   & $2.22$~\tablenotemark{j} & 22:29:31.82 & +47:42:24.8 & $173.34^\circ$ & $2.58 \times 10^{-7}$& 17.6 \\
     $\sigma$ Canis Majoris & $0.513 \pm 0.108$  & $2.78$~\tablenotemark{f}  & 07:01:43.15 & -27:56:05.4  & $19.24^\circ$ & $9.39 \times 10^{-6}$ & 0.376 \\
     VV Cephei & $0.599 \pm 0.083$  & $5.42$~\tablenotemark{j}  & 21:56:39.14 & +63:37:32.0 & $174.34^\circ$ & $2.58 \times 10^{-7}$ & 5.13  \\
     $\theta$ Delphini & $0.629 \pm 0.029$   & $0.870$~\tablenotemark{f}  & 20:38:43.99 & +13:18:54.4 & $157.15^\circ$ & $2.58 \times 10^{-7}$ & 29.0 \\
     V381 Cephei & $0.631 \pm 0.086$   & $1.92$~\tablenotemark{j}  & 21:19:15.69 & +58:37:24.6 & $174.68^\circ$ & $2.58 \times 10^{-7}$ & 13.0\\
      V424 Lacertae &  $0.634 \pm 0.075$   & $1.91$~\tablenotemark{k} & 22:56:26.00 & +49:44:00.8 & $173.30^\circ$ & $2.58 \times 10^{-7}$ & 13.0 \\
     HR 861 & $0.639 \pm 0.039$    & $1.10$~\tablenotemark{l}   & 02:56:24.65  & +64:19:56.8  & $176.43^\circ$ & $2.58 \times 10^{-7}$ & 22.2  \\ 
     145 Canis Major &  $0.697 \pm 0.078$    & $1.68$~\tablenotemark{h}   & 07:16:36.83 & -23:18:56.1 & $10.26^\circ$ & $3.99 \times 10^{-5}$ & 0.079\\
     V809 Cassiopeia &  $0.730 \pm 0.074$   & $2.85$~\tablenotemark{m}  & 23:19:23.77 & +62:44:23.2 & $178.76^\circ$ & $2.58 \times 10^{-7}$ & 6.56 \\
     HR 8248 & $0.746 \pm 0.039$  & $0.675$~\tablenotemark{l} & 21:33:17.88 & +45:51:14.5 & $176.30^\circ$ & $2.58 \times 10^{-7}$ & 26.6 \\
     Deneb/$\alpha$ Cygni & $0.802 \pm 0.066$  & $1.41$~\tablenotemark{n} & 20:41:25.9 & +45:16:49.0 & $178.15^\circ$ & $2.58 \times 10^{-7}$ & 11.0 \\
\enddata
\tablecomments{
$^a$\cite{lambdavelorum},
$^b$\cite{antares},
$^c$\cite{epsilonpegasi},
$^d$\cite{betelgeuse},
$^e$\cite{qcar},
$^f$\cite{zetacephei_xicygni_nspuppis_o1canismajoris_12pegasi},
$^g$\cite{rigel1,rigel2},
$^h$\cite{wcar},
$^i$\cite{cetauri},
$^j$\cite{5lacertae},
$^k$\cite{v424lacertae},
$^l$\cite{hr861},
$^m$\cite{v809cassiopeia},
$^n$\cite{deneb}.
}
\end{deluxetable*}


\begin{acknowledgements}
We thank Christopher Fontes from Los Alamos National Laboratory for his help in determining the monochromatic opacity for 0.511 MeV $\gamma-$rays. We are grateful to Regina Caputo, Yong-Zhong Qian and Yudai Suwa for useful discussions. We also thank Carolyn A. Kierans for providing us with the effective area for COSI. CL acknowledges support from the NSF grant PHY-2309973, and from the National Astronomical Observatory of Japan, where part of this work was conducted. M.\,M. is supported by NSF Grant No. AST-2108466. M.\,M. also acknowledges support from the 
Institute for Gravitation and the Cosmos (IGC) Postdoctoral Fellowship. FXT and EF acknowledge support from  NSF under grant 2154339 entitled "Neutrino Emission From Stars".
\end{acknowledgements}

\bibliography{refs.bib}{}
\bibliographystyle{aasjournal}

\end{document}